\documentclass[prl,amssymb,amsmath,preprintnumbers,showpacs,twocolumn]{revtex4}

\usepackage{graphicx}
\usepackage{color}
\usepackage{verbatim}
\usepackage{subfigure}
\include{subcaption}
\usepackage{amssymb}

\usepackage{amsmath}
\usepackage{setspace} 

\newcommand{\PRA}{Phys. Rev. A}

\newcommand{\PRE}{Phys. Rev. E}
\newcommand{\PRL}{Phys. Rev. Lett.}

\newcommand{\NJP}{New. J. Phys.}

\begin{document}

\title{Fast forward to the classical adiabatic invariant}

\author{Christopher Jarzynski}
\affiliation{Institute for Physical Science and Technology, University of Maryland, College Park, MD 20742 USA}
\affiliation{Department of Chemistry and Biochemistry, University of Maryland, College Park, MD 20742 USA}
\affiliation{Department of Physics, University of Maryland, College Park, MD 20742 USA}
\author{Sebastian Deffner}
\affiliation{Theoretical Division and Center for Nonlinear Studies, Los Alamos National Laboratory, Los Alamos, NM 87545, USA}
\affiliation{Department of Physics, University of Maryland Baltimore County, Baltimore, MD 21250 USA}
\author{Ayoti Patra}
\affiliation{Department of Physics, University of Maryland, College Park, MD 20742 USA}
\author{Yi\u{g}it Suba\c{s}\i}
\affiliation{Department of Chemistry and Biochemistry, University of Maryland, College Park, MD 20742 USA}
\affiliation{Theoretical Division, Los Alamos National Laboratory, Los Alamos, NM 87545, USA}

\date{\today}

\begin{abstract}
We show how the classical action, an adiabatic invariant, can be preserved under non-adiabatic conditions.
Specifically, for a time-dependent Hamiltonian $H = p^2/2m + U(q,t)$ in one degree of freedom, and for an arbitrary choice of action $I_0$, we construct a ``fast-forward'' potential energy function $V_{\rm FF}(q,t)$ that, when added to $H$, guides all trajectories with initial action $I_0$ to end with  the same value of action.
We use this result to construct a local dynamical invariant $J(q,p,t)$ whose value remains constant along these trajectories.
We illustrate our results with numerical simulations. Finally, we sketch how our classical results may be used to design approximate quantum shortcuts to adiabaticity.
\end{abstract}


\maketitle

For a classical system in one degree of freedom, the action variable $I= \oint p\cdot {\rm d}q$ is an adiabatic invariant~\cite{goldstein_1980}.
As an example, when the length of a pendulum is slowly varied, both its energy $E$ and frequency of oscillation $\omega$ change with time, but their ratio $E/\omega$, which is proportional to the action, remains constant.

In this paper we consider a system whose Hamiltonian $H(q,p,t) = p^2/2m + U(q,t)$ varies at an arbitrary rate, hence the action $I(q,p,t)$ does not remain constant:
if at time $t=0$ we launch a collection of trajectories, each with the same initial action $I_0$, then at later times their actions will generally differ from one another, and from the initial action.
But suppose we want these trajectories  to ``return home'' at a specified later time $\tau$, i.e.\ we demand that the action of each trajectory be equal to $I_0$ at $t=\tau$, given that its action had this value at $t=0$.
In this paper we solve for the additional forces that are required to steer the trajectories back to the action $I_0$ at $t=\tau$.
More precisely, we show how to construct an auxiliary ``fast-forward'' potential $V_{\rm FF}(q,t)$ with the following property.
Under the dynamics generated by the Hamiltonian $H_{\rm FF} = H + V_{\rm FF}$, {\it all trajectories that begin with action $I_0$ at $t=0$ will end the same action, $I_0$, at $t=\tau$}.
Throughout this paper, the action $I(q,p,t)$ is defined with respect to the original Hamiltonian $H(q,p,t)$.

We were led to this topic through our interest in quantum {\it shortcuts to adiabaticity}~\cite{torrontegui_2013},
and (as we briefly discuss later) we expect our results will prove useful in the design of such shortcuts for guiding a quantum system to a desired energy eigenstate.
The primary focus of this paper, however, is a self-contained problem of general theoretical interest in elementary classical dynamics, for which we obtain a simple and appealing solution (Eq.~\ref{eq:solution}).

Consider a classical system in one degree of freedom, described by a kinetic-plus-potential Hamiltonian
\begin{equation}
\label{eq:Hdef}
H(z,t) = \frac{p^2}{2m} + U(q,t) \quad,\quad z = (q,p)
\end{equation}
$H$ varies with time during the interval $0\le t\le\tau$, but is constant outside this interval.
We assume that $H$ is twice continuously differentiable with respect to time~\cite{arnold_1978}, hence both $\partial H/\partial t$ and $\partial^2 H/\partial t^2$ vanish at $t=0$ and $t=\tau$.
In the Supplemental Material (SM), we discuss how this assumption can be relaxed.

The term {\it energy shell} will denote a level curve of $H(z,t)$, that is the set of all points where $H$ takes on a particular value, $E$, at time $t$. 
We will assume that each energy shell forms a simple, closed loop in phase space.
The function
\begin{equation}
\Omega(E,t) = \int {\rm d}z \, \theta\left[ E - H(z,t) \right] = \oint_E p\cdot {\rm d}q
\end{equation}
is the volume of phase space enclosed by the energy shell $E$ of $H(z,t)$, and the {\it action},
\begin{equation}
\label{eq:Idef}
I(z,t) = \Omega(H(z,t),t) ,
\end{equation}
is the volume enclosed by the energy shell that contains the point $z$.
Eq.~\ref{eq:Idef} implies
\begin{equation}
\label{eq:PB}
\{I, H \} \equiv \frac{\partial I}{\partial q} \frac{\partial H}{\partial p} - \frac{\partial I}{\partial p} \frac{\partial H}{\partial q} = 0
\end{equation}

Let us choose an arbitrary action value $I_0>0$, and define the {\it adiabatic energy} $\bar E(t)$ by the condition
\begin{equation}
\label{eq:adiabaticEnergy}
\Omega(\bar E(t),t) = I_0
\end{equation}
The {\it adiabatic energy shell} ${\cal E}(t) = \{ z \vert H(z,t) = \bar E(t) \}$ is the level curve of $H(z,t)$ with the value $\bar E(t)$, enclosing a phase space volume $I_0$.
Hence $I(z,t) = I_0$ for all $z \in {\cal E}(t)$.

At $t=0$, the adiabatic energy shell ${\cal E}(0)$ defines a set of initial conditions that form a closed loop in phase space.
As trajectories evolve under $H(z,t)$ from these initial conditions, this loop evolves in time,
\begin{equation}
{\cal L}(t) = \{ z = z_t(z_0) \vert z_0 \in {\cal E}(0) \}
\end{equation}
where $z_t(z_0)$ indicates the trajectory that evolves under $H(z,t)$ from initial conditions $z_0$.
If $H$ varies slowly with time, then these trajectories remain close to the adiabatic energy shell, but under more general conditions the loop ${\cal L}(t)$ strays away from ${\cal E}(t)$ for $t>0$.

Now consider an auxiliary potential $V_{\rm FF}(q,t)$, let $z_t^{\rm FF}(z_0)$ indicate evolution under $H_{\rm FF} = H+V_{\rm FF}$, and consider the loop
\begin{equation}
\label{eq:LFF}
{\cal L}_{\rm FF}(t) = \{ z = z_t^{\rm FF}(z_0) \vert z_0 \in {\cal E}(0) \}
\end{equation}
that evolves under $H_{\rm FF}$ from the initial conditions defined by ${\cal E}(0)$.
Our aim is to construct $V_{\rm FF}(q,t)$ such that ${\cal L}_{\rm FF}(\tau) = {\cal E}(\tau)$: we want the auxiliary potential to guide trajectories faithfully back to the adiabatic energy shell at the final time $t=\tau$.
The notation FF, for ``fast-forward''~\cite{masuda_2010,masuda_2011,torrontegui_2012,torrontegui_2012njp,masuda_2014,kiely_2015,kazutaka_2015,deffner_2016}, indicates that $V_{\rm FF}$ drives the system rapidly to a destination that it would otherwise have reached during a slow process.
We now describe how to construct a potential $V_{\rm FF}(q,t)$ with this property.

\begin{figure}[tbp]
\includegraphics[trim = 1in 2.5in 0in 2in , scale=0.25,angle=0]{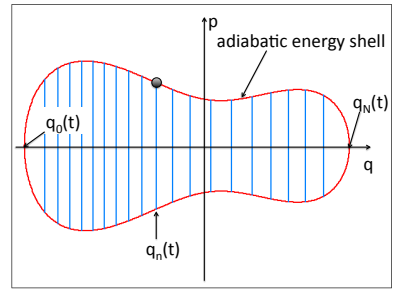}
\caption{
The region of phase space enclosed by the adiabatic energy shell ${\cal E}(t)$ is divided by line segments at $\{ q_n(t) \}$ into vertical strips of equal phase space volume.
The motion of these lines is described by velocity and acceleration fields $v(q,t)$ and $a(q,t)$.
}
\label{fig:strips}
\end{figure}

Imagine a set of line segments at locations $q_1(t), \cdots q_{N-1}(t)$, that divide the region of phase space enclosed by the adiabatic energy shell ${\cal E}(t)$ into $N\gg 1$ narrow strips of equal phase space volume; see Fig.~\ref{fig:strips}.
Let $q_0(t)$ and $q_N(t)$ denote the left and right turning points of the adiabatic energy shell.
In the limit $N\rightarrow\infty$, the time-dependence of these line segments defines a velocity field $v(q,t)$ and an acceleration field $a(q,t)$:
\begin{equation}
\label{eq:qdotqddot}
\frac{{\rm d} q_n}{{\rm d}t} = v(q_n,t) \quad,\quad \frac{{\rm d}^2 q_n}{{\rm d}t^2} = a(q_n,t) = \frac{\partial v}{\partial q} v + \frac{\partial v}{\partial t}
\end{equation}
Since $\partial H/\partial t=\partial^2 H/\partial t^2=0$ at $t=0$ and $t=\tau$ (see comments following Eq.~\ref{eq:Hdef}) we have
\begin{equation}
\label{eq:atrest}
v(q,0) = v(q,\tau) = 0
\quad,\quad
a(q,0) = a(q,\tau) = 0
\end{equation}
We claim that the desired fast-forward potential satisfies
\begin{subequations}
\label{eq:solution}
\begin{equation}
-\frac{\partial V_{\rm FF}}{\partial q} = ma
\end{equation}
therefore it is given by~\footnote{
The choice of setting the lower bound of integration at $q_0(t)$ is arbitrary.
A different choice would modify $V_{\rm FF}(q,t)$ by an additive function $\phi(t)$ having no effect on the dynamics.
}
\begin{equation}
\label{eq:Vff}
V_{\rm FF}(q,t) = - \int_{q_0(t)}^q {\rm d}q^\prime \, ma(q^\prime,t)
\end{equation}
\end{subequations}
By Eq.~\ref{eq:atrest}, $V_{\rm FF}(q,t)$ vanishes at the start and end of the process.
Since $v(q,t)$ and $a(q,t)$ depend on the value $I_0$, different choices of $I_0$ generally produce different fast-forward potentials $V_{\rm FF}(q,t)$.
We now show that an auxiliary potential given by Eq.~\ref{eq:solution} will indeed produce the desired result, for an arbitrary but fixed $I_0$. 

We begin by solving for the velocity field $v(q,t)$.
The volume of the region of phase space that is enclosed by the energy shell, and is located to the left of a point $q \in [q_0,q_N]$, is given by
\begin{equation}
\label{eq:Sdef}
S(q,t) = 2 \int_{q_0(t)}^q {\rm d}q^\prime \, \bar p(q^\prime,t)
\end{equation}
where $\bar p(q,t) = \left[ 2m(\bar E - U) \right]^{1/2}$ specifies the upper branch of the adiabatic energy shell.
By construction, $v(q_n(t),t)$ is the velocity of a line segment $q_n(t)$ that evolves at fixed $S$: $({\rm d}/{\rm d}t) S(q_n(t),t) = 0$.
Hence
\begin{equation}
v(q,t) = \frac{{\rm d}q}{{\rm d}t} \biggr\vert_S = - \frac{\partial_t S}{\partial_q S}
\end{equation}

Now consider a point in phase space, $(q_n(t),p_n(t))$, attached to the top of the $n$'th line segment: $p_n = \bar p(q_n,t)$ (see Fig.~\ref{fig:strips}).
As the shape of the energy shell and the locations of the line segments vary parametrically with time, this point $(q_n,p_n)$ moves in phase space, ``surfing" the upper branch of the energy shell.
This motion is described by the equations
\begin{equation}
\label{eq:surfer}
\dot q_n = v(q_n,t) \qquad,\qquad \dot p_n =  -p_n v^\prime(q_n,t)
\end{equation}
where the equation for $\dot p_n$ is obtained by demanding that the phase space volume of the strip between neighboring vertical lines, $\delta S_n \equiv 2 p_n (q_{n+1}-q_n)$, remain constant.
Eq.~\ref{eq:surfer} also describes the motion of a point attached to the bottom of one of the vertical lines.
We easily verify that Eq.~\ref{eq:surfer} is generated by a Hamiltonian
\begin{equation}
\label{eq:Kdef}
K(q,p,t) = p v(q,t)
\end{equation}
Therefore if we start with initial conditions distributed over the energy shell ${\cal E}(0)$, and we evolve trajectories from these initial conditions under the Hamiltonian $K(q,p,t)$, then these trajectories cling to the evolving adiabatic energy shell, with each trajectory attached to the upper or lower end of one of the vertical line segments.
Hence the flow generated by $K$ {\it preserves the adiabatic energy shell}, in the following sense: for each time step $\delta t$, this flow maps points on ${\cal E}(t)$ to points on ${\cal E}(t+\delta t)$.
Equivalently, the action $I(z,t)$ is conserved under this flow, for those trajectories with action $I_0$:
\begin{equation}
\label{eq:dIdt=0}
0 = \frac{\partial I}{\partial t} + \frac{\partial I}{\partial q} \dot q + \frac{\partial I}{\partial p} \dot p
= \frac{\partial I}{\partial t} + \{ I,K \} \qquad \forall \, z \in {\cal E}(t)
\end{equation}

Next, we construct a Hamiltonian $G(z,t) \equiv H + K$, which generates equations of motion
\begin{equation}
\label{eq:ztildedot}
\dot q = \frac{p}{m} + v(q,t) \quad , \quad
\dot p = -U^\prime(q,t) - p v^\prime(q,t)
\end{equation}
Along a trajectory $z(t)$ obeying these dynamics,
\begin{equation}
\label{eq:Idot}
\dot I = \frac{\rm d}{{\rm d}t} I(z(t),t) = \frac{\partial I}{\partial t} + \{ I , H\} + \{ I,K \}
\end{equation}
Eqs.~\ref{eq:PB}, \ref{eq:dIdt=0} and \ref{eq:Idot} imply that $\dot I = 0$ for all $z\in{\cal E}(t)$.
Thus the flow generated by $G = H + K$ preserves the adiabatic energy shell.
This is easily understood: with each time step $\delta t$, the term $K(z,t)$ generates a flow that maps ${\cal E}(t)$ onto ${\cal E}(t+\delta t)$ while the term $H(z,t)$ generates flow parallel to the adiabatic energy shell.
As a consistency check, we can verify directly from Hamilton's equations that the flow generated by $G$ preserves the adiabatic energy shell (see SM).

To this point, we have constructed a Hamiltonian $G = H+K$ that generates trajectories which cling to the adiabatic energy shell ${\cal E}(t)$.
Along these trajectories, $I(z,t)$ remains constant. 
We now introduce a change of variables that effectively transforms $K(z,t)$ into the potential energy function $V_{\rm FF}(q,t)$ that we seek.

Consider the evolution of the observables
\begin{equation}
\label{eq:QPdef}
Q = q \qquad,\qquad P = p + mv(q,t)
\end{equation}
along a trajectory that evolves under Eq.~\ref{eq:ztildedot}.
By direct substitution we get
\begin{equation}
\label{eq:Zdot}
\frac{{\rm d}Q}{{\rm d} t} = \frac{P}{m} \qquad,\qquad \frac{{\rm d}P}{{\rm d} t} = -U^\prime(Q,t) + ma(Q,t)
\end{equation}
using Eq.~\ref{eq:qdotqddot}.
Eq.~\ref{eq:Zdot} is generated by the Hamiltonian
\begin{equation}
H_{\rm FF}(Z,t) = H(Z,t) + V_{\rm FF}(Q,t)
\end{equation}
where $Z=(Q,P)$ and $V_{\rm FF}$ satisfies Eq.~\ref{eq:solution}.
Thus Eq.~\ref{eq:QPdef} defines a time-dependent transformation ${\cal M}_t:z\rightarrow Z$, which maps any trajectory $z(t)$ evolving under $G(z,t)$ to a counterpart trajectory $Z(t)$ evolving under $H_{\rm FF}(Z,t)$.
Now consider specifically a trajectory $z(t)$ that evolves, under $G$, from initial conditions on the adiabatic energy shell ${\cal E}(0)$.
As we have already seen, this trajectory remains on the adiabatic energy shell ${\cal E}(t)$ for all times $t\in [0,\tau]$.
Under the mapping ${\cal M}_t$, its image $Z(t)$ (which evolves under $H_{\rm FF}$) is displaced along the momentum axis by an amount $mv(q,t)$ (Eq.~\ref{eq:QPdef}).
By Eq.~\ref{eq:atrest}, $Z(t)$ begins and ends on the adiabatic energy shell: $Z(0) \in {\cal E}(0)$, $Z(\tau) \in {\cal E}(\tau)$.
This is precisely the behavior we desired to generate, which concludes our proof.

In Eq.~\ref{eq:LFF} we used ${\cal L}_{\rm FF}(t)$ to denote a loop in phase space evolving under $H_{\rm FF}$.
The results of the previous paragraph can be written compactly as follows:
\begin{equation}
\label{eq:compact}
{\cal M}_t: {\cal E}(t) \rightarrow {\cal L}_{\rm FF}(t)
\end{equation}
At any time $t$, ${\cal L}_{\rm FF}(t)$ is the image of ${\cal E}(t)$ under the transformation defined by Eq.~\ref{eq:QPdef} (see Fig.~\ref{fig:intermediateTime} below).
This result implies that the function $J(q,p,t) \equiv I(q,p-mv(q,t),t)$ is a {\it local dynamical invariant}.
That is, if a trajectory $z(t)$ is launched from the energy shell ${\cal E}(0)$ and then evolves under $H_{\rm FF}$, then the value of $J$ is conserved along this trajectory: $J(z(t),t) = I_0$.
For consistency, we can verify directly from Hamilton's equations that ${\rm d}J/{\rm d}t=0$ for any point $z\in{\cal L}_{\rm FF}$ (see SM).

\begin{figure}[tbp]
   \subfigure{
   \label{fig:initialConditions}
   \includegraphics[trim = 2in 5in 0in 0in, scale=0.30]{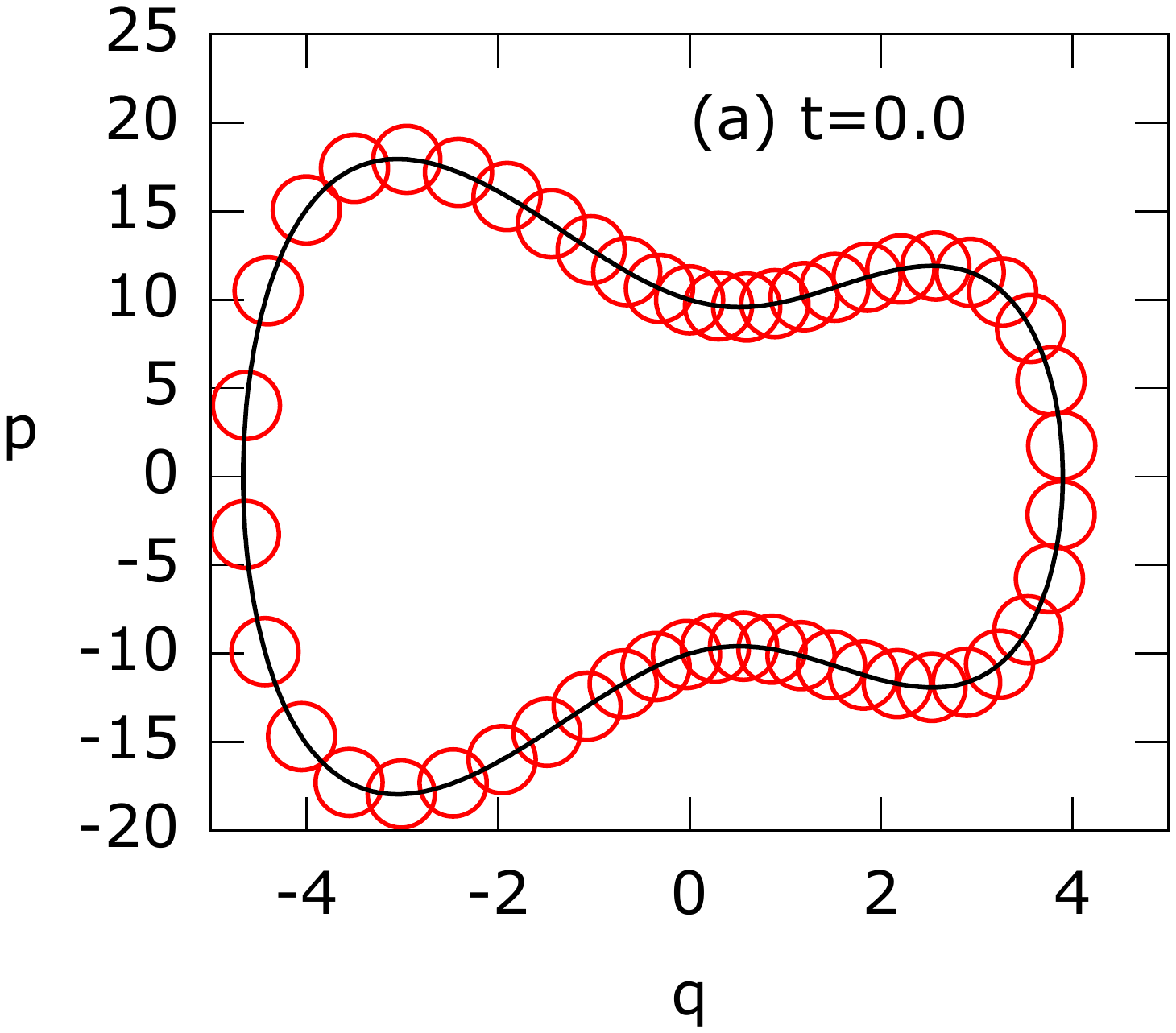}
   }
   \subfigure{
   \label{fig:finalConditions_noCD}
   \includegraphics[trim = 2in 5in 0in 0in, scale=0.30]{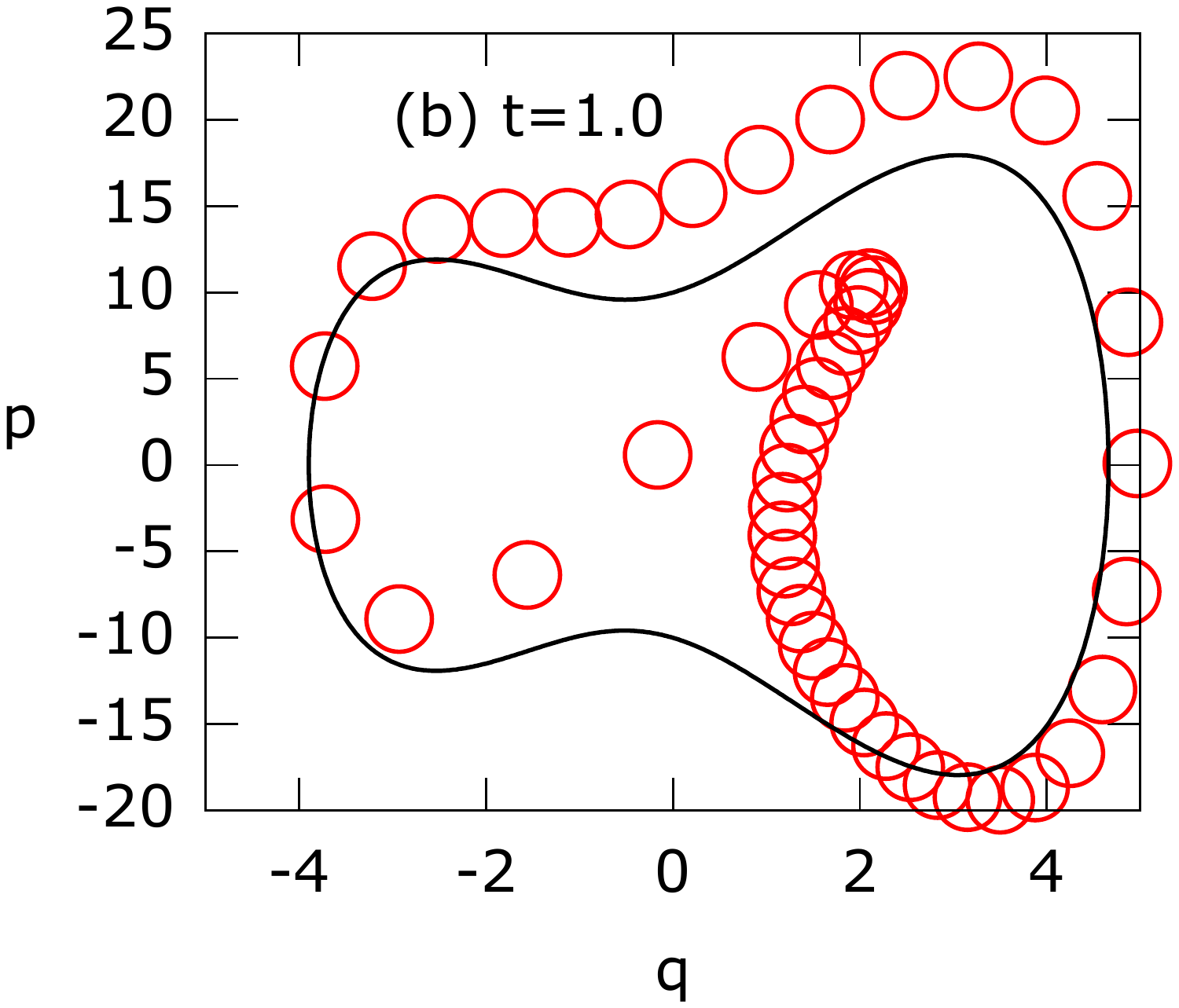}
   }
   \subfigure{
   \label{fig:finalConditions_CD}
   \includegraphics[trim = 2in 5in 0in 0in, scale=0.30]{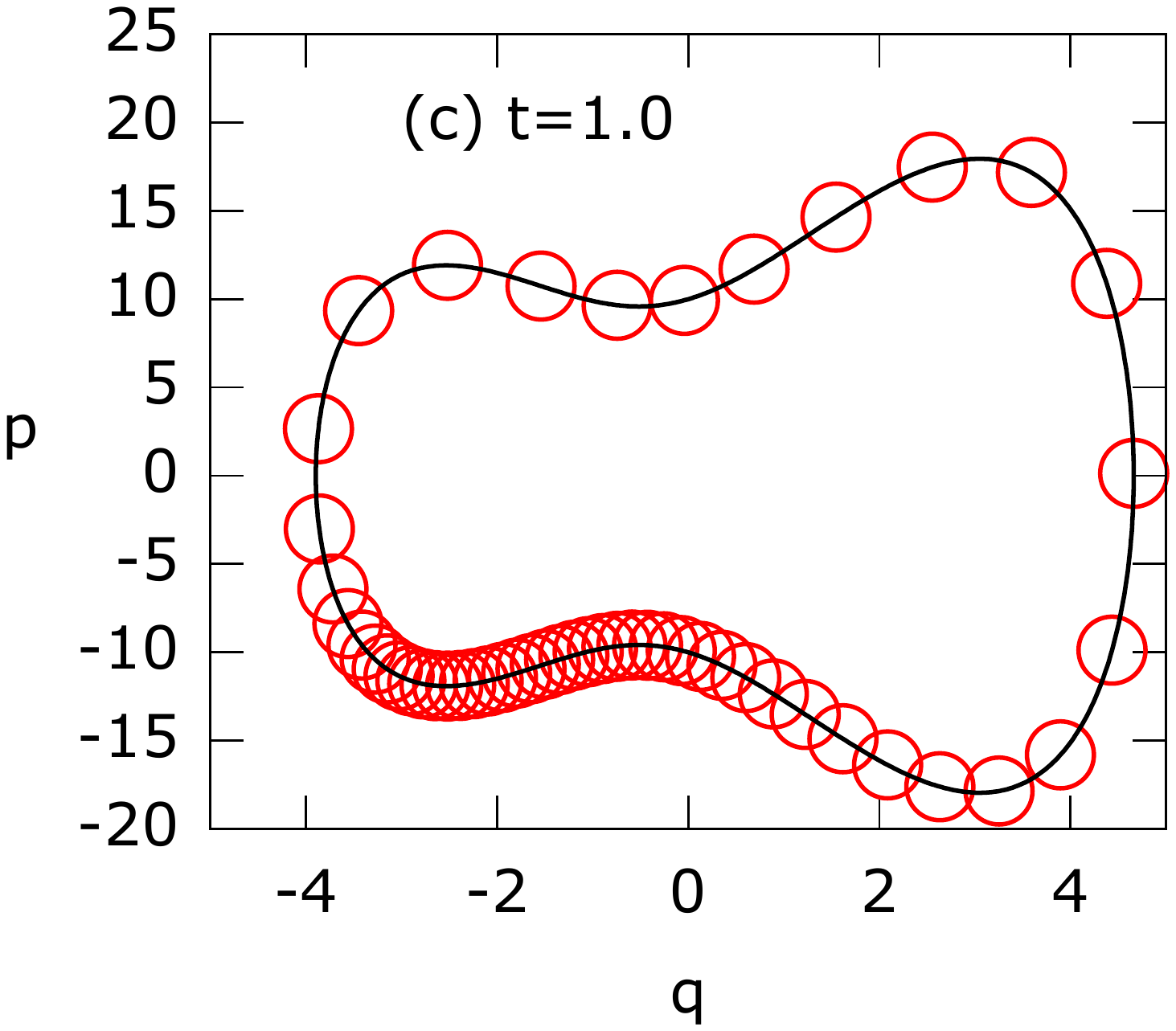}
   }
\caption{
Initial (a) and final (b,c) conditions for trajectories launched from a single energy shell ${\cal E}(0)$.
The trajectories in panel (b) evolved under $H(z,t)$ (Eq.~\ref{eq:modelHamiltonian}), while those in panel (c) evolved under $H_{\rm FF} = H+V_{\rm FF}$.
The solid black curves show the adiabatic energy shell ${\cal E}(t)$ at initial and final times.
}
\label{fig:proofOfPrinciple}
\end{figure}

To illustrate our results, we chose the dimensionless Hamiltonian
\begin{equation}
\label{eq:modelHamiltonian}
H(z,t) = \frac{p^2}{2} + q^4 - 16 q^2 + \alpha(t) q
\end{equation}
with $\alpha(t) = 4\cos(\pi t/\tau)[5- \cos(2\pi t/\tau)]$ and $\tau = 1.0$.
We considered an initial adiabatic energy shell ${\cal E}(0)$ with energy $\bar E(0) = 50.0$ and action $I_0 = 214.035$.
We numerically determined the fields $v(q,t)$ and $a(q,t)$ and constructed $V_{\rm FF}(q,t)$ according to Eq.~\ref{eq:solution}.
We then generated 50 initial conditions on the energy shell ${\cal E}(0)$, shown in Fig.~\ref{fig:initialConditions}, and we performed two sets of simulations.
In the first set, trajectories were evolved from these initial conditions under $H(z,t)$.
In the second set, trajectories were evolved from the same initial conditions under the Hamiltonian $H_{\rm FF} = H+V_{\rm FF}$.
In the absence of the fast-forward potential $V_{\rm FF}$, the trajectories belonging to the first set have final actions $I(z,\tau)$ that span a range of values, as seen in Fig.~\ref{fig:finalConditions_noCD}.
By contrast, the addition of $V_{\rm FF}$ guides the second set of trajectories back to the adiabatic energy shell ${\cal E}(\tau)$, where each trajectory ends with $I(z,\tau)=I_0$; see Fig.~\ref{fig:finalConditions_CD}.
Note, however, that while the initial conditions in Fig.~\ref{fig:initialConditions} are spaced uniformly with respect to the microcanonical measure, this is not the case for the final conditions in Fig.~\ref{fig:finalConditions_CD}.
As discussed in the SM, this non-uniformity is due to the fact that $V_{\rm FF}(q,t)$ depends on the choice of $I_0$.

We also performed simulations with a shorter duration, $\tau=0.2$.
After constructing $V_{\rm FF}(q,t)$ for this faster protocol, we simulated 50 trajectories evolving under $H_{\rm FF} = H+V_{\rm FF}$, using the initial conditions in Fig.~\ref{fig:initialConditions}.
Fig.~\ref{fig:intermediateTime} depicts a snapshot of these trajectories at $t=\tau/2$.
The two closed curves show the adiabatic energy shell ${\cal E}(t)$ and its image under the mapping $p \rightarrow p+v(q,t)$ {(see Eq.~\ref{eq:QPdef}).
This figure confirms Eq.~\ref{eq:compact}: the trajectories evolving under $H_{\rm FF} = H+V_{\rm FF}$ are located on a loop ${\cal L}_{\rm FF}(t)$ that is obtained by ``shearing'' the instantaneous energy shell ${\cal E}(t)$ along the momentum axis, by an amount $mv(q,t)$.

\begin{figure}[tbp]
\includegraphics[trim = 2in 5in 0in 0in, scale=0.30]{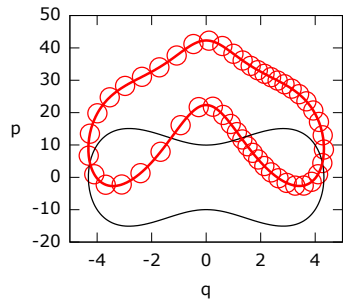}
\caption{
A snapshot, at $t=\tau/2$, of 50 trajectories evolving under $H_{\rm FF}(z,t)$ using a rapid protocol (see text).
The closed black loop is the adiabatic energy shell ${\cal E}(t)$, and the red loop above it is constructed by displacing each point on the lower loop by an amount $mv(q,t)$ along the $p$-axis.
As predicted by Eq.~\ref{eq:compact}, the trajectories coincide with the red loop.
}
\label{fig:intermediateTime}
\end{figure}

For so-called {\it scale-invariant driving}~\cite{def14}, the time-dependence of $U(q,t)$ is described by scaling and translation parameters $\gamma(t)$ and $f(t)$:
$U(q,t) = (1/\gamma^2) U_0[(q-f)/\gamma]$.
We then obtain $v(q,t) = (\dot\gamma/\gamma)(q-f)+\dot f$ and
\begin{equation}
V_{\rm FF}(q,t) = -\frac{m}{2} \frac{\ddot\gamma}{\gamma}(q-f)^2 - m\ddot f q
\end{equation}
which does not depend on $I_0$.~\cite{def14}
In this rather special case, every trajectory evolving under $H_{\rm FF}$ returns to its adiabatic energy shell at $t=\tau$, $J(z,t)$ is a {\it global} dynamical invariant -- it is the Lewis-Riesenfeld invariant~\cite{lewis_1968,lewis_1969a} -- and microcanonical initial distributions are mapped to microcanonical final distributions.

The problem that we have studied has a quantum analogue, introduced by Masuda and Nakamura~\cite{masuda_2010,masuda_2011}:
given $\hat H(t) = -(\hbar^2/2m) (\partial^2/\partial q^2) + U(q,t)$,
construct $V_{\rm FF}^{(n)}(q,t)$ such that evolution under $\hat H + V_{\rm FF}^{(n)}$ causes a selected eigenstate $\varphi_n(q,0) \equiv \langle q \vert n(0)\rangle$ of $\hat H(0)$ to evolve to the corresponding eigenstate $\varphi_n(q,\tau)$ of $\hat H(\tau)$.
This problem has been solved for both Schr\" odinger~\cite{masuda_2010,masuda_2011,torrontegui_2012,torrontegui_2012njp,masuda_2014,kiely_2015,kazutaka_2015} and Dirac~\cite{deffner_2016} dynamics, but the solution generically becomes singular at the nodes of $\varphi_n(q,t)$ (see e.g.\ Eq.\ 5 of Ref.~\cite{torrontegui_2012}), 
hence a well-behaved $V_{\rm FF}^{(n)}(q,t)$ cannot generally be constructed for $n>0$.\footnote{
The special case of scale-invariant driving is an exception to this statement.}
Our result offers an alternative approach:
for the classical Hamiltonian $H(z,t) = p^2/2m + U(q,t)$, construct $V_{\rm FF}^{(n)}(q,t)$ as the fast-forward potential corresponding to the action $I_0 = 2\pi\hbar [n+(1/2)]$.
This potential is free from singularities, and for large $n$ the Correspondence Principle suggests that evolution under $\hat H + V_{\rm FF}^{(n)}$ will cause the initial wavefunction $\varphi_n(q,0)$ to evolve approximately to the final wave function $\varphi_n(q,\tau)$.
Preliminary numerical results support this expectation~\cite{pat16}.

It is also interesting to compare our analysis with the {\it counter-diabatic} approach, where the quantum eigenstate $\vert n(t)\rangle$~\cite{demirplak_rice_2003,berry_2009} or the classical action $I(z,t)$~\cite{jarzynski_2013,deng_2013,patra_2016} is preserved along the entire trajectory.
In the classical case this is achieved at the cost of adding a momentum-dependent term $H_{\rm CD}(z,t)$ rather than a potential $V_{\rm FF}(q,t)$, to the Hamiltonian.
For scale-invariant driving~\cite{def14}, $H_{\rm CD}$ coincides with our term $K$ (Eq.~\ref{eq:Kdef}), but more generally $H_{\rm CD}$ is a nonlinear function of both $q$ and $p$, which may be complicated~\cite{patra_2016} and difficult to implement experimentally.

It is natural to ask whether our results can be applied to systems with $d>1$ degrees of freedom.
In certain situations of experimental relevance, such as ultracold gases in optical lattices, a separation of variables reduces a three-dimensional problem to an effectively one-dimensional one~\cite{bloch_2005,masuda_2014}, providing a potential platform to test our predictions.
More generally, the distinction between integrable, chaotic, and mixed phase space systems becomes crucial for $d$-dimensional systems~\cite{berry_1978}.
For integrable systems, the transformation to action-angle variables~\cite{goldstein_1980} may provide a useful first step to extending our results, but for mixed or chaotic systems the task is likely to be more challenging.

Adiabatic invariants enjoy a distinguished history in quantum and classical mechanics~\cite{navarro_2006},
but the problem of how to achieve adiabatic invariance under non-adiabatic conditions has gained attention only recently.
Here we have shown how to construct a potential $V_{\rm FF}(q,t)$ that guides trajectories launched from a given energy shell of an initial Hamiltonian to the corresponding energy shell of the final Hamiltonian,
so that the initial and final values of action are identical for every trajectory.

We acknowledge financial support from the U.S. National Science Foundation under grant DMR-1506969 (CJ), the U.S. Department of Energy through a LANL DirectorÕs Funded Fellowship (SD), and the U.S. Army Research Office under contract number W911NF-13-1-0390 (AP, YS).


%

\end{document}